\documentclass[preprint,journal]{vgtc}       
\pdfoutput=1 

\usepackage{mathptmx}
\usepackage{graphicx}
\usepackage{times}
\usepackage{epstopdf}
\usepackage{textcomp}
\usepackage{cite}
\usepackage{listings}

\usepackage[bookmarks,backref=true,linkcolor=blue]{hyperref} 
\hypersetup{
  pdfauthor = {},
  pdftitle = {},
  pdfsubject = {},
  pdfkeywords = {},
  colorlinks=true,
  linkcolor= blue,
  citecolor= blue,
  pageanchor=true,
  urlcolor = blue,
  plainpages = false,
  linktocpage
}

\onlineid{0}

\vgtccategory{Research}

\vgtcinsertpkg

\preprinttext{Mississippi State University, Department of Computer Science and Engineering, Technical Report MSU-140331-01}

\title{Efficacy of Images Versus Data Buffers: Optimizing Interactive Applications Utilizing OpenCL for Scientific Visualization}
\author{Donald W. Johnson and T. J. Jankun-Kelly}
\authorfooter{Mail Stop 9637, Mississippi State, MS 39762, USA. E-mail: \texttt{tjk@acm.org}
}

\shortauthortitle{Donald W. Johnson \MakeLowercase{\textit{et al.}}: Efficacy of Images Versus Data Buffers: Optimizing Interactive Applications Utilizing OpenCL for Scientific Visualization}

\abstract{This paper examines an algorithm using dual OpenCL image buffers to optimize data streaming for ensemble processing and visualization. Image buffers were utilized because they allow cached memory access, unlike simple data buffers, which are more commonly used. OpenCL image object performance was improved by allowing upload and mapping into one buffer to occur concurrently with mapping and/or processing of data in another buffer. This technique was applied in an interactive application allowing multiple flood extent maps to be combined into a single image, and allowing users to vary input image sets in real time. The efficiency of this technique was tested by varying both dimensions of input images and number of iterations; computation scaled linearly with number of input images, with best results achieved using ~4k images. Tests were performed to determine the rate at which data could be moved from data buffers to image buffers, examining a large range of possible image buffer dimensions. Additional tests examined kernel runtimes with different image and buffer variants. Limitations of the algorithm and possible applications are discussed. 

} 


 \nocopyrightspace

\begin{document}

\maketitle

\section{Introduction}

Modern GPUs are powerful tools utilized for volume rendering, scientific modeling, medical imaging and other tasks involving parallel computation. As data transfer between the CPU and GPU is slow, the use of GPU memory, or \emph{buffers}, is vital for performant analysis and visualization. Though data buffers are more commonly used in visualization, this work uses OpenCL image buffers primarily for the processing and depiction of data (in this case, a flood visual analytics system). Image buffers offer cached access to GPU memory at the cost of imaging coding on upload; this work outlines approaches to diminish this cost while demonstrating the benefits of image buffers. 

Basic interaction with a GPU has three main steps: uploading data to the device, processing or display of uploaded data, and retrieval of results.  Because both upload and download are significantly slower than the potential processing speed of a GPU, performance is best when those steps are performed only once. This can only be done if the dataset to be processed can fit entirely into graphics memory. When this is not the case, or when the size of data can not be known beforehand, two strategies are commonly employed. Data can be sampled to provide lower resolution information that will fit in available graphics memory; alternatively, when the goal of processing is computation instead of display, it may be necessary to process the input data at its original density. In this case, data must be subdivided into pieces that will fit into available graphics memory. There are at least two ways to attempt this subdivision. Data can be spatially subdivided, or broken into segments, that can be processed discreetly. However, if steps are not taken to synchronize the processing of boundaries, this approach can cause artifacts at the edges of segments. Alternatively, to avoid subdividing data sets, processing may be done iteratively, building a working composite which is continuously updated as layers of data are processed. If the scope of the data is sufficiently vast, it may be necessary to apply both spatial and iterative subdivision to process the data set. Thus, intelligent management of the CPU--GPU transfer of data is paramount. 

OpenCL is the only cross platform, hardware-independent, high-level graphics hardware programming interface widely available. When using OpenCL to move data to or from the GPU, two main formats are available, image buffers (henceforth referred to as images) and data buffers (henceforth referred to as buffers). While buffers are simple linear memory structures, images have an internal format dependent on the GPU driver, normally a block-based structure.  For data processing, images have several advantages over buffers. The primary advantage is that image reads are cached \cite{NVIDIA},  allowing an order of magnitude faster access than data that resides in GPU main memory. Images are also easily shared with the graphics environment for display, and support accelerated packing and unpacking of data. However, despite these advantages, current literature focuses primarily on data transfer between data buffers, and seldom discusses methods of efficiently streaming data into image objects. This work explores the application of image buffers to data processing and visualization.

The  contributions of this work are:
\begin{itemize}
	\item Determination of the amount of acceleration achievable by using double buffering techniques to hide data transfer and transformation when using OpenCL images.
	\item Recommendations for maximum uncompressed data size that can be processed by the dual buffer algorithm while maintaining an interactive frame rate ($  \ge $ 10 FPS).
	\item Determination of the relationship between image dimensions and image transformation speed.
	\item Determination of the acceleration resulting from replacing buffers with images in the tested kernels.
\end{itemize}

\section{Motivation}

\begin {figure}[tb]
	\label{fig:program} 
	\includegraphics[scale=0.40]{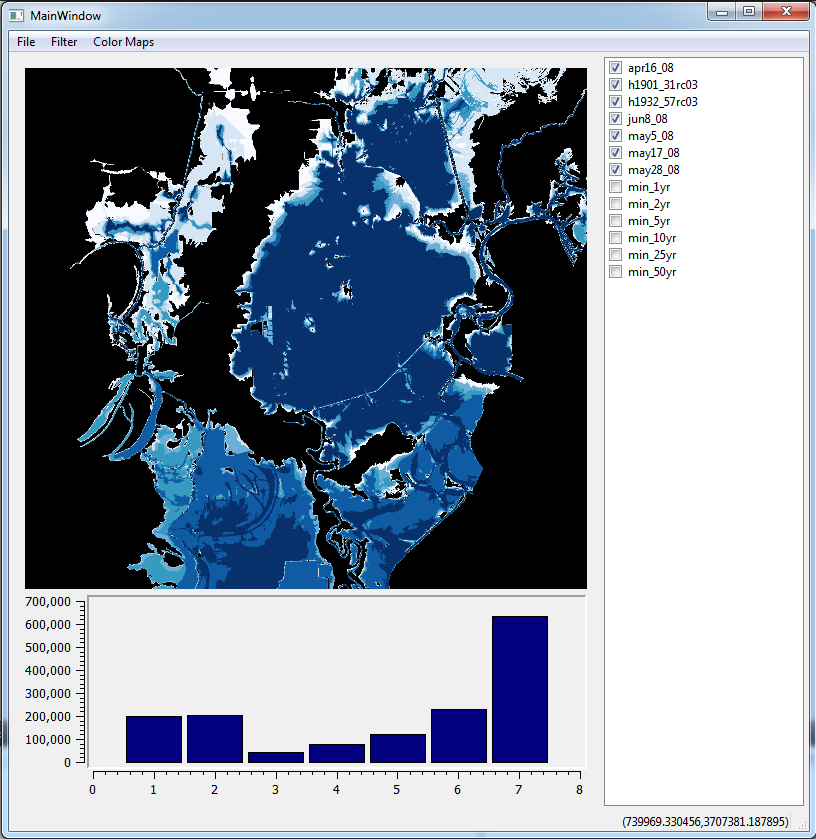}
\caption{Flood extent analysis program. The flood analytics program highlights region of multiple flood overlap using a blue saturation map (high overlap corresponds to high saturation); these extents (and others) are calculated from the $N$ input surfaces into a single composite image using the discussed technique. The list in the right panel allowed users to add or remove surfaces from consideration in real time. A composite image of the currently selected images is shown in the upper left panel. The lower left panel displays a histogram showing the frequency of overlap classes, indicating distribution of flood occurrences.}
\end{figure}

The algorithm discussed in this work was created as part of a solution designed to perform ensemble processing and visualization of flood coverage images (Figure \ref{fig:program}). Given a collection of simulated and observed flood data, the system allow users to rapidly find regions of high and low flood overlap. A second goal was to identify  image clusters, which, within the scope of the project, were defined as subsets of input data where all members had a significant degree of overlap over the entire data. The analysis program was required to handle extremely large input images, particularly if LIDAR (LIght Detection and Ranging) datasets were used. In addition there was no hard limit to the number of images that might be considered simultaneously. The need the process an unknown number of potentially large images required a solution capable of handling input sets so vast that all data would be unable to fit in GPU memory. However CPU-based analysis would not be fast enough to provide real time interaction with the large datasets. Coupled with a requirement that users be able to frequently and quickly change the working set of images, a streaming OpenCL approach was utilized. 

The flood visualization system consists of several OpenCL kernels that perform the outlier and clustering calculations whose results are displayed in OpenGL. While the kernels are specific to this particular domain, the approach can be applied in other systems with coupled GPU-powered analysis and visualization. Initial results using a single receiving image were slower than desired, eventually leading to the development of the dual buffer algorithm herein discussed next.

\section{Releated Work}

Most early examples of graphics hardware being used to aid computation are in the area of volume rendering. One early example is Cabral et al.~\cite{Cabral:1994} where graphics hardware was used to increase the performance of back-projection based volume rendering; hardware accelerated results were more than 100 times faster than the CPU. Lum et al.~\cite{Lum:2001} illustrated how graphics hardware and parallel rendering could be combined to allow visualization of large time varying datasets. Hadwiger et al.~\cite{Hadwiger} later illustrated another method whereby graphics hardware could be used to accelerate volume rendering through the inclusion of a segment volume used to isolate individual objects in the volume. Eventually frameworks that simplified access to hardware acceleration and parallel rendering began to appear (e.g., Bhaniramka and Demange~\cite{Bhaniramka:2002}). Slightly more recent work includes a summary of techniques usable in real time volume rendering \cite{Engel:2004} and advanced illumination methods for volume rendering \cite{Hadwiger:2008}. Recent work in ray casting techniques (most of the early volume rendering was back projected ray casting) includes Lux and Fr\"{o}hlich~\cite{Lux:2011} and Zhu et al. \cite{Zhu:2012}. Early works use texture GPU memory in a manner that presages the algorithm described here, though our work uses general purpose image buffers; our data is also streamed. 

As the aforementioned methods were being developed, graphics hardware continued to advance. The fixed functionality pipeline of early graphics hardware was replaced by programmable units. These units first had to be controlled with assembly code; for example NVIDIA's language cg \cite{NVCG}. From assembly languages, the programing of graphics hardware progressed to high-level graphics programing languages with syntax modeled on C, namely Microsoft's High Level Shader Language (HLSL) \cite{HLSL}, and OpenGL's OpenGL Shader Language (GLSL) \cite{GLSL}. For general computation purposes, both of these languages would be overtaken by new languages specifically for this purpose (all previous languages were designed for graphics but could be forced to do general computation). The primary languages in this category are NVIDIA's Cuda \cite{Cuda} and the Khronos Groups' OpenCL (Open Compute Language) \cite{khronos}.

 In the area of optimizing data streaming, Vo et al.~\cite{Vo:2010} introduces a framework supporting multi-core systems. Unfortunately, the system described does not interact with GPUs, largely due to limitations of the connected graphics framework (VTK).

Basic information on recommended usage for OpenCL and CUDA can be found in \cite{AMD} for AMD GPUs and \cite{NVIDIA} for NVIDIA GPUs. There are several notable studies on streamed data processing with CUDA. In their work on the Dax toolkit \cite{Dax:2011}, Moreland et al. present a high level framework capable of reorder task to minimize, or in some cases eliminate, overhead from I/O. In "CudaDMA"\cite{CudaDMA}, Bauer et al. describe a framework that optimizes stream performance in CUDA by using warp specialization techniques and support for different types of buffing models. Another framework (ISP) and study is presented by Ha et al. \cite{ISP:2012}; this study included analysis of different streaming modes, and the effects of reordering upload, execution, download, and optionally compression of data. Another framework with capabilities for streaming data between differnt types of processing units is presented by Vo et al. in \cite{Hyperflow:2012}. A recent work of Sewell et al. \cite{Sewell:2013} presented a framework supporting use of CUDA capable graphics hardware for parallel visualization. Another recent work in this area is Rosen \cite{Rosen:2013}, which describes a system for visualizing memory conflicts generated when running CUDA kernels. Such conflicts, caused by hardware dependent variables, greatly slow the performance of computation, and must be checked for and solved on a per-device basis. Studies using OpenCL are more limited. One important study using OpenCL was done Spafford et al \cite{Maestro:2010}, which studied the effects of buffering techniques, work group size, and data transfer size, when using OpenCL buffers.

Our work extends available work by evaluating the effectiveness of using images (as opposed to buffers) in the use of visualization OpenCL kernels. We also study the effect of the non-linear cost of moving data from buffers to images, especially has image size increases. Finally, a comparison of the relative run times of kernels that differ only in data structure (images or buffers) is conducted. These tests together provide the necessary information to choose appropriate memory transfer structures.

\section{Algorithm}

\begin {figure}[tb]
	\includegraphics[scale=0.35]{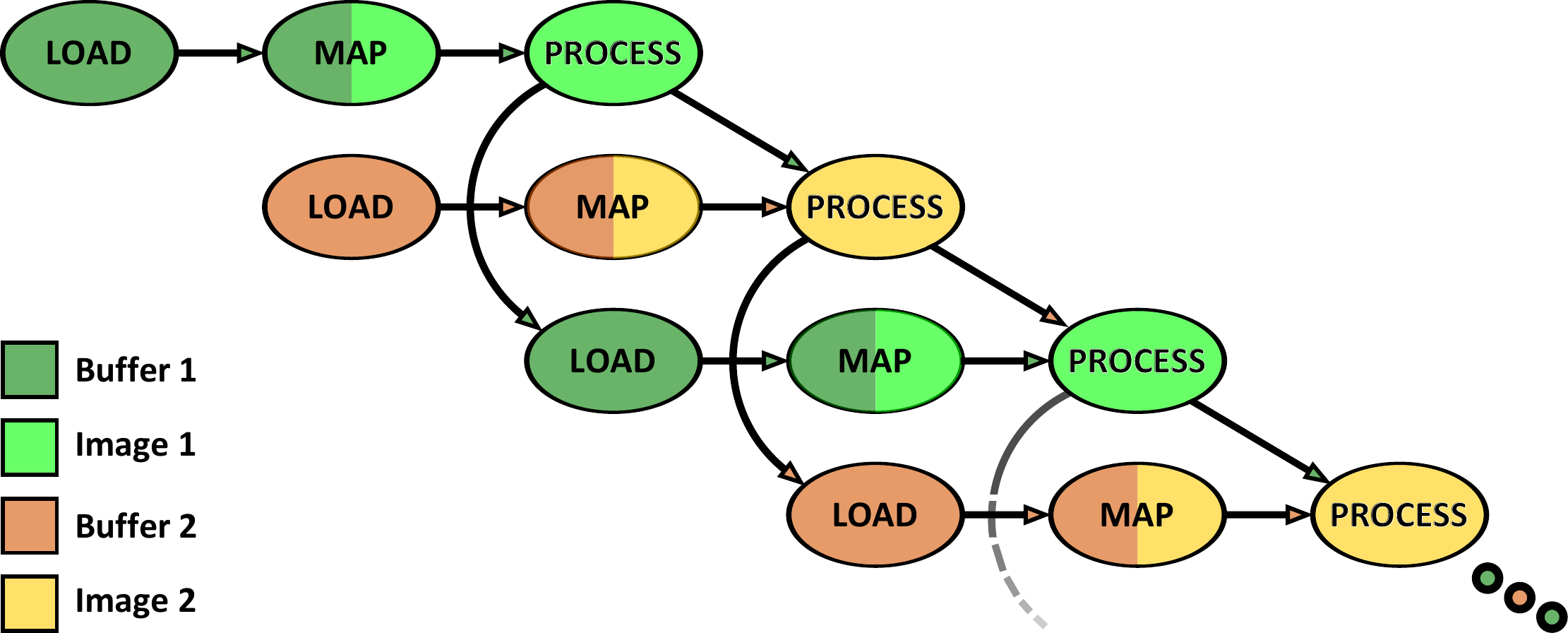}
	\caption{Dual Buffer Algorithm Flow: Order of operations for the final dual buffer algorithm (2b final) with two input buffer/image pairs. Arrows indicate dependency: an event will not start until all events connecting to it have completed.}
	\label{fig:depend}
\end{figure}

A naive approach to streaming data involves the use of a single pair of buffers to transfer data between the client and device. Pseudo code illustrating this method is shown in Figure \ref{fig:Alg1}. The algorithm herein presented exploits the ability of the GPU to simultaneously transfer and process data.  Pseudo code illustrating this method is shown in Figure \ref{fig:Alg2}. When a single pair of buffers or images is utilized, the GPU will, at best, cycle between uploading and processing data. However, by utilizing two pairs of either images or buffers to send and receive data,  processing may take place in one pair while I/O takes place in the other. The flow of operations and their dependencies for the dual buffer method are illustrated in Figure \ref{fig:depend}.

\begin {figure}[tb]
	\begin{enumerate}
		\item Clear accumulation Buffers
		\item set the values of \textbf{pos1}, \textbf{pos2}, and \textbf{i} to \textbf{0},  \textbf{1}, and \textbf{0}
		\item Load data into \textbf{input\_i} from client memory
		\item Run accumulation kernel where
			\begin{itemize}
				\item \textbf{input\_i} is an input
				\item \textbf{accum\_[pos1]} is an input
				\item \textbf{accum\_[pos2]}i s an output
			\end{itemize}
		\item swap the values \textbf{pos1} and \textbf{pos2}
		\item increment the value of \textbf{i}
		\item if \textbf{i $\ge$ N} stop, otherwise go to 3 \\
	\end{enumerate}
	\vspace{-5 mm}
	\tiny Where \textbf{i}, \textbf{pos1}, and \textbf{pos2} are integers \\
	\textbf{input\_i}, \textbf{accum\_i[0]}, and \textbf{accum\_i[1]} are OpenCL images \\
	\textbf{N} is the number of images to processes 
	\caption{Naive Approach -- 1 Buffer}
	\label{fig:Alg1}
\end{figure}

\begin {figure}[tb]
	\begin{enumerate}
		\item Clear accumulation Buffers
		\item set the values of \textbf{pos1}, \textbf{pos2}, and \textbf{i} to \textbf{0},  \textbf{1}, and \textbf{0}
		\item Load data into \textbf{input\_i[pos1]} from client memory
		\item Run accumulation kernel where
			\begin{itemize}
				\item \textbf{input\_i[pos1]} is an input
				\item \textbf{accum\_[pos1]} is an input
				\item \textbf{accum\_[pos2]}i s an output
			\end{itemize}
		\item swap the values \textbf{pos1} and \textbf{pos2}
		\item increment the value of \textbf{i}
		\item if \textbf{i $\ge$ N} stop, otherwise go to 3 \\
	\end{enumerate}

	\vspace{-5 mm}
	\tiny Where \textbf{i}, \textbf{pos1}, and \textbf{pos2} are integers \\
	\textbf{input\_i[0]}, \textbf{input\_i[1]}, \textbf{accum\_i[0]}, and \textbf{accum\_i[1]} are OpenCL images \\
	\textbf{N} is the number of images to processes 
	\caption{Dual Buffer Approach }
	\label{fig:Alg2}
\end{figure}

For either the naive single buffer approach or the proposed dual buffer approach, multiple implementation methods are possible. In the course of testing the efficiency of the dual buffer approach a total of three implementation methods were tested. 

The first method, which seemed to be the obvious approach, utilized \textbf{clWriteImage()} to transfer data directly from client memory to permanent device-resident image structures. The control algorithm (\emph{1b initial}) employed a single device-resident image, while the experimental algorithm (\emph{2b initial}) utilized two device-resident image structures. Unfortunately, the function \textbf{clWriteImage()} first creates a hidden duplicate linear-format copy of the data in client memory, and secondly a linear-format copy on the device side, before finally transforming the data into image-format on the device side. This resulted in a transfer rate on the test system that was only about 40\% efficient. 

A second implementation method that eliminated the unneeded data copy in client memory by using data buffers initially before converting to images was also tested. The limitation of this approach is that data transfer and transformation (into an image format) was coupled together, preventing other memory transfers from being initiated until after the final transformation of the data had completed. Due to this bottleneck, we focus our study on the naive approach and the optimized one discussed next. 

The third and final implementation method decoupled the transfer and transformation of data. This was accomplished by creating permanent dedicated receiving buffers on the device, which were paired with permanent dedicated image structures. The control algorithm (\emph{1b final}) utilized one buffer-image pair, while the experimental algorithm (\emph{2b final}) utilized two buffer-image pairs. Just as with the second implementation method, data was initially stored in OpenCL buffers allocated in client memory. Data was then transfered from client memory to the receiving buffers as one operation using \textbf{clCopyBuffer()}. A second operation, \textbf{clCopyBufferToImage()}, was utilized to handle the transformation of data into image format once the transfer completed. The advantage of this system is that, the entire time data in one device buffer is being transformed into an image, as well as the time taken for the resulting image to be processed, becomes a window in which data can be transfered from the client into the second buffer-image pair.  

All  described methods were designed for use with an asynchronous OpenCL queue. While it is possible to utilize these methods in synchronous mode, efficiency will be poor. OpenCL event references were used to coordinate tasks within each variant algorithm. It is important to ensure that a kernel does not execute utilizing as input either an image that was currently being updated, or an image that had already been processed and not yet been updated. Likewise, event references were utilized to ensure that new data was not loaded into an image or buffer that was currently being employed by either a kernel or a buffer-to-image copy.

\section{Testing}

\begin {figure*} [tb]
	\begin{center}
	\begin{tabular}{ l  r  r   r  r  r  r }
	\hline \\ [-2ex]
	& & \multicolumn{4}{c}{Processing Time  ($\mu$s)} \\
	\\ [-2.5ex] \hline \\ [-1.5ex]
	 Data Set & N & 1b initial & 2b initial & 1b final & 2b final \\
	\\ [-2.5ex] \hline \\ [-1.5ex]
	\texttildelow2k & 10 & - & - & 16,901.01	 & 10,060.57 \\
	 & 100 &- & -  & 154,798.84 & 97,795.6 \\
	 & 1,000 & - & - & 1,544,108.34	 & 967,545.27 \\
	 & 10,000 &- & -  & 15,253,992.48 & 9,641,951.45 \\
	\\ [-2.5ex] \hline \\ [-1.5ex]
	\texttildelow4k & 10 & 79,224.52 & 69,423.91 & 47,112.72 & 35,602.03 \\
	 & 100 & 779,224.59 & 702,200.18  & 414,403.68 & 338,179.28 \\
	 & 1,000 & 7,913,592.64 & 6,936,346.74  & 4,160,507.91	 & 3,354,251.81 \\
	 & 10,000 & 78,868,351.01 & 70,314,931.82  & 41,207,446.97 & 33,782,002.28 \\
	\\ [-2.5ex] \hline \\ [-1.5ex]
	\texttildelow8k & 10 &- & - & 185,410.59 & 179,550.34 \\
	 & 100 &- & -  & 1,649,694.40 & 1,778,611.77 \\
	 & 1,000 & - & - & 16,571,037.83 & 17,769,066.44 \\
	 & 10,000 &- & -  & 165,437,132.54 & 178,148,199.43 \\
	\hline
	\end{tabular}
	\end{center}
	\caption{Algorithm Run Times: This table shows the processing time for all recorded tests. The initial naive algorithms (1b initial, 2b initial) were only tested with the \texttildelow4k image set, whereas the optimized algorithms (1b final, 2b final) were also tested with the \texttildelow2k and \texttildelow8k datasets. For each combination of dataset and algorithm, time values increase in a linear pattern, with the value for $N$=10 being slightly higher than the trend indicated by the other points. Additionally, note that for all tests the dual buffer algorithm performed faster, except when using \texttildelow8k images.  Because the dual buffer algorithm uses slightly more resources, this degradation of performance may be caused by resource contention.}
	\label{fig:4ktable}
\end{figure*}

Tests were performed in three different areas. The first tests measured the impact, on the overall runtime of an accumulation kernel, of using either single or dual buffer algorithms for data transfer. The second tests determined how buffer-to-image copying performance changed depending on target image dimensions. The third set of tests examined the runtime effects of using either images or buffers in computation kernels which performed the same calculations. 

All tests were performed on a machine running Windows 7 Sp1 with 8 GB of installed RAM and a single 1080p monitor. The system utilized a single graphics card, an AMD Radeon HD 7950. Tests were performed with default settings and using the Catalyst 13.4 driver. 
\subsection{Dual Buffer Tests}

Timing data was gathered by recording the total time necessary to transmit each of N images from client to device memory, and process the received images with an accumulation kernel. Tests  were conducted for N equals 10, 100, 1,000, and 10,000. In all cases, timing began immediately after the accumulation image-structures used by the kernel were cleared and ended as soon as the final call to the accumulation kernel was reported complete by the OpenCL runtime. Actual time values were provided by the function \textbf{gettimeofday()} implemented in Windows using \textbf{GetSystemTimeAsFileTime()}. All recorded times were the result of averaging 100 repeated tests of the same number of iterations. 

Timing tests were initially performed only on images with dimensions of ~4k (3712x4416). After these tests had been concluded, the optimized algorithms (1b final, 2b final) were tested again with images measuring ~2k (1856x2208) and ~8k (7424x8832). This was done to see if image dimensions influenced processing time. Original plans also called for an additional test image set measuring ~16k (14848x17664), however the graphics hardware utilized did not support images of this size.

\subsection{Buffer to Image Transformation Tests}

Testing of the single and dual buffer algorithms revealed that the time required to transform data from buffers to images in OpenCL was not linearly associated with image size. Ideally, in order to determine the buffer-to-image performance for each possible image dimension, the entire range of possible image dimensions would be tested. However, the domain of possible image dimensions is too large for exhaustive testing, with most modern graphics hardware allowing for over 268 million possible image dimensions. Therefore, two sampling grids were utilized to test  buffer-to-image transfer rate for different image dimensions. The first sampling pattern began with a 128x128 pixel image and increased image dimensions in steps of 128 pixels, culminating in an 16,384x16,384 image and sampling 16,384 possible images dimensions. The second sampling pattern started with a 100x100 image and increased image dimensions in steps of 100 to a maximum image size of 16,300x16,300, resulting in 26,569 samples. Two sampling patterns were utilized to increase confidence that any resulting pattern was not simply a result of a given sampling pattern. For each image dimension, time required to transform data from an appropriately sized buffer to the image was measured. Each measurement was repeated 1,000 times, and the average transformation time was recorded. 

\subsection{Buffer vs Image Kernel Runtime Tests}

Kernel runtimes were measured for 4 different computational kernels. The first tested kernel (image-kernel 1) was the kernel from the previous single and dual buffer tests which utilized an image for input and a pair of images for storage of intermediate results. The second tested kernel (buffer-kernel 1) was an accumulation kernel which utilized a buffer for input and another buffer for storage of intermediate results. It was possible to utilize a single buffer because one may both read-from and write-to the same buffer within a kernel, as opposed to images, which must be either read-only or write-only within a kernel (in current versions of OpenCL). The second pair of kernels were designed to eliminate differences between image-kernel 1 and buffer-kernel 1. The third tested kernel (image-kernel 2) was an image-based accumulation kernel where input image dimensions are passed as kernel arguments rather than obtained by querying the input image. The purpose of this modification was to allow image-based kernel behavior to more closely match buffer-based kernel behavior (since there is no query support for buffers). The final tested kernel (buffer-kernel 2) was a buffer-based accumulation kernel that used two buffers to handle intermediate results, in order to create a buffer-based kernel that generated the same number of memory read and write operations as the image-based kernels. Each of the four tested computation kernels was run 10,000 times. Performance measurements were taken using the profiling tools available through AMD's CodeXL program.

\section{Results and Discussion}

\subsection{Dual Buffer Results}
\begin{figure}[!t]
	\includegraphics[scale=0.45]{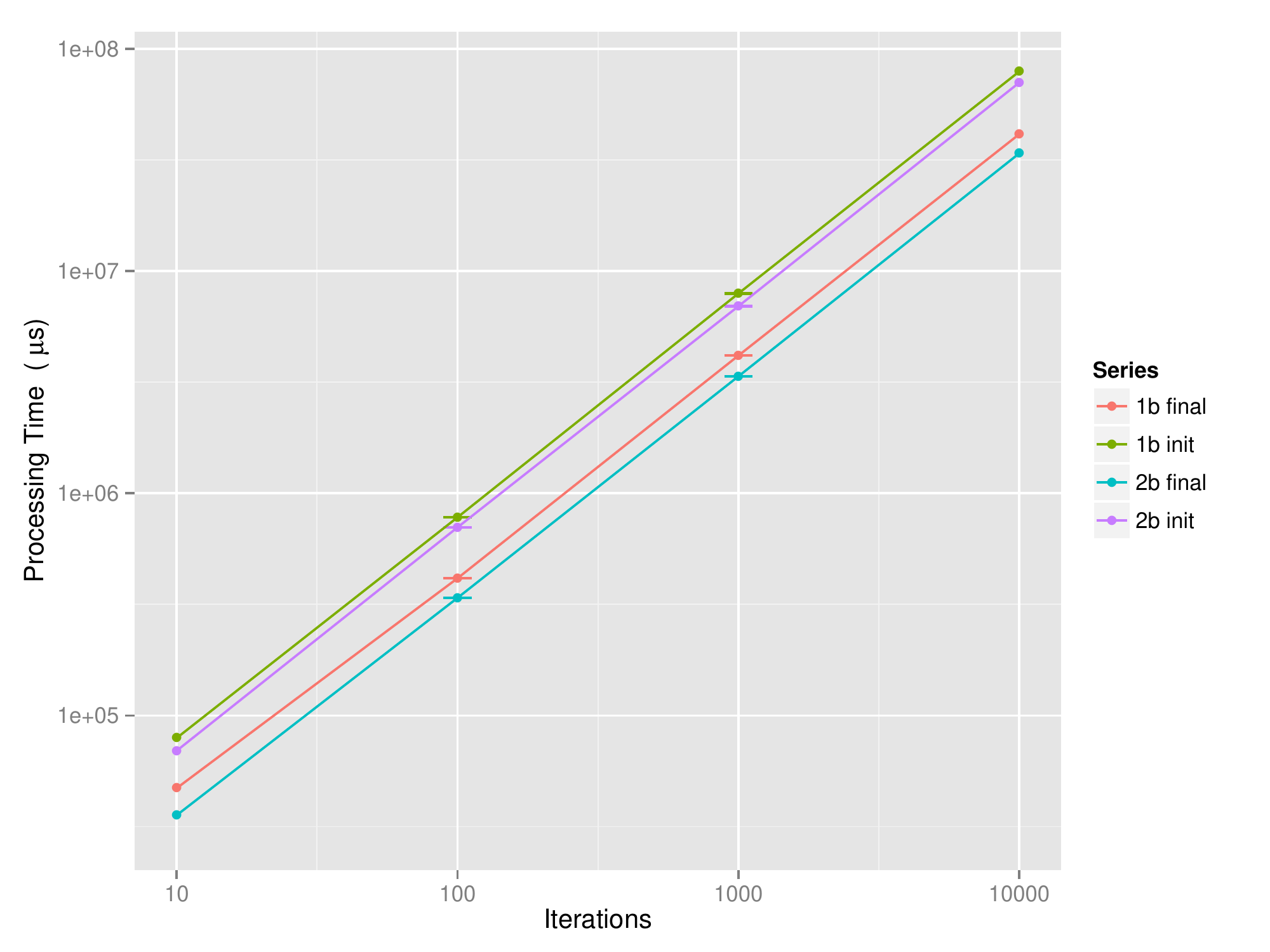}
	\caption{Image Processing Times for 4k Images: Although the lines in this figure appear parallel, they are not. Lines with different slopes appear parallel in logarithmic space because space itself is distorted. The difference between the lines at bottom of the graph is orders or magnitude less than the difference between the lines at the top.   } 
	\label{fig:graph_sp4k}  
\end{figure}

Timing data was collected for the described algorithms when run with ~4k image inputs; for simplicity, this paper only reports timing data for the initial naive algorithms (1b initial, 2b initial) and the final optimized algorithms (1b final, 2b final). In addition, timing data was only recorded for the optimized algorithms (1b final, 2b final) when run with ~2k and ~8k datasets. Raw timing data for the ~4k data set can be seen in Fig \ref{fig:4ktable} and Fig \ref{fig:graph_sp4k}. 

Testing of the first two algorithms (1b initial, 2b initial) showed that the two buffer approach was significantly faster. However, when transfer rates were calculated (see Fig \ref{fig:4ktrans}) it was clear that neither technique was near the theoretical transfer rate cap (8.0 GB/s for a PCIE 2.0 bus). The outlook for the dual buffer technique was somewhat improved by the discovery that available bandwidth to the device on the test system, as reported by AMD's bandwidth test program (BufferBandwidth.exe), was actually only 5.07 GB/s. Even so, neither technique appeared to be effective. 

\begin{figure*}	
	\includegraphics[scale=0.80]{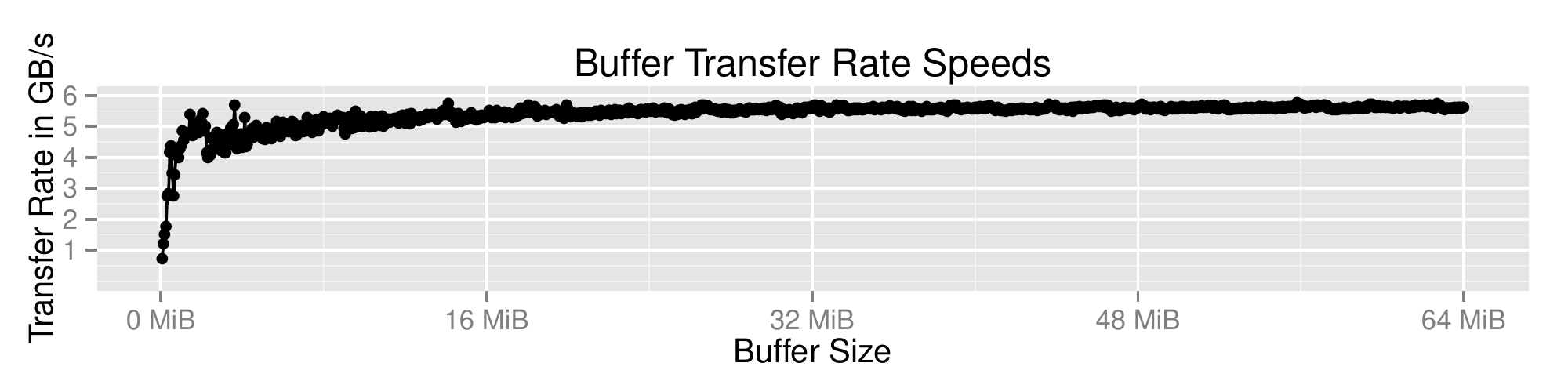}  

	\caption{Transfer Rate of a Copy from CPU Resident Buffer to a GPU Resident Buffer as Function of Buffer Size: buffer transfer rate is very poor with small buffer sizes, but improves rapidly and settles on a rough average of 5.1 GB/s. The values from this test were used as the baseline values when calculating the efficiency in other tests.  }
	\label{fig:graph_transfer} 
\end{figure*}

For accurate calculation of efficiency, for each algorithm, with each input set, transfer speed tests were done for all buffer sizes between 64 kiB and 64 MiB in intervals of 64 kiB. The results of these test can bee seen in Fig  \ref{fig:graph_transfer}. For all image sizes, efficiency was calculated by dividing the time to transfer and processes data with a given algorithm and dataset, by the buffer-to-buffer transfer time for a matching amount data with an identically sized buffer.

\begin {figure} [!tb]
	\begin{tabular}{ l  r  r  r  }
	\hline
	 Alg & Iterations & Transfer Rate   & Efficiency  \\
		&     N &   GB/s & \% \\
	\\ [-2.5ex] \hline \\ [-1.5ex]
	1b init & 10 & 2.069 & 39.15 \\
	 & 100 & 2.104 & 39.81 \\
	 & 1,000 & 2.071 & 39.19 \\
	 & 10,000 & 2.078 & 39.33 \\
	\\ [-2.5ex] \hline \\ [-1.5ex]
	2b init & 10 & 2.361 & 44.68 \\
	 & 100 & 2.334 & 44.17 \\
	 & 1,000 & 2.363 & 44.72 \\
	 & 10,000 & 2.331 & 44.11\\
	\\ [-2.5ex] \hline \\ [-1.5ex]
	1b final & 10 & 3.479 & 65.84 \\
	 & 100 & 3.956 & 74.85 \\
	 & 1,000 & 3.940 & 74.55 \\
	 & 10,000 & 3.978 & 75.27 \\
	\\ [-2.5ex] \hline \\ [-1.5ex]
	2b final & 10 & 4.604 & 87.12 \\
	 & 100 & 4.847 & 91.72  \\
	 & 1,000 & 4.887 & 92.47 \\
	 & 10,000 & 4.852 & 91.82\\
	\\ [-2.5ex] \hline \\ [-1.5ex]
	\end{tabular}
	\caption{4k Image Transfer Rate and Efficiency: efficiency was calculated by comparing the time to transfer and process data, with the indicated algorithms, to the time required for a simple buffer-to-buffer transfer of the same amount of data.}
	\label{fig:4ktrans}
	 
\end{figure}

\begin{figure}[!t]
	\includegraphics[scale=0.45]{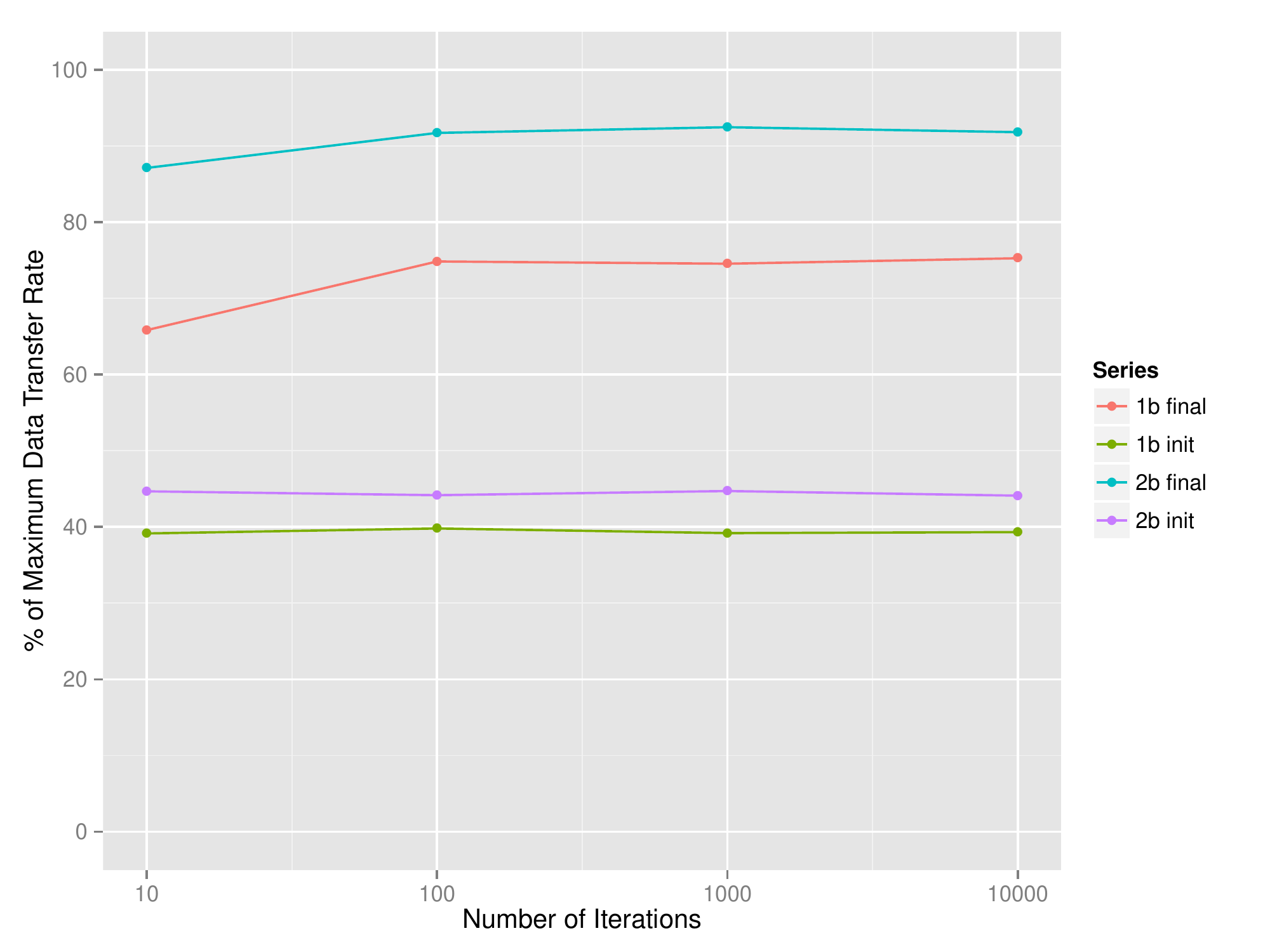}
	\caption{Efficiency for 4k Images: The small loss of efficiency when $N=10$ is likely due to warm up time for the GPU.  }
	\label{fig:graph_e4k}    
\end{figure}

Both one buffer and two buffer techniques showed significant improvement when optimized. The efficiency of the one buffer technique showed dramatic improvement, increasing by 68\% to 91\%, dependent on iterations (N). The efficiency of the two buffer technique also improved dramatically, increasing by 94\% to 108\%, dependent on iterations (N).

For both optimized algorithms (1b final, 2b final) the slower per image processing time for the $N=10$ tests can be attributed to GPU warm up time (see Figure \ref{fig:graph_e4k}). The unoptimized algorithms (1b initial, 2b initial) do not exhibit this behavior, but their low efficiency overall makes it likely that the GPU never even reached full transfer speed while running. For all tests, running time appears to grow linearly with increasing iterations (N), excepting $N=10$ where speed is slightly slower than a linear pattern would predict.

Timing data for \texttildelow2k and \texttildelow8k tests can be seen in Fig \ref{fig:4ktable}; transfer rate and efficiency numbers are shown in Fig \ref{fig:2ktrans}. In the case of \texttildelow2k images, transfer rates are universally worse than for the \texttildelow4k images. Some of the change in data transfer rate is the result of less efficient buffer copying when using smaller buffers. In terms of efficiency, the dual buffer algorithm (2b final) performs almost as well at \texttildelow2k resolution as at \texttildelow4k; however, surprisingly, the one buffer algorithm performed significantly worse at \texttildelow2k then at \texttildelow4k. 

\begin{figure}[!t]
	\includegraphics[scale=0.45]{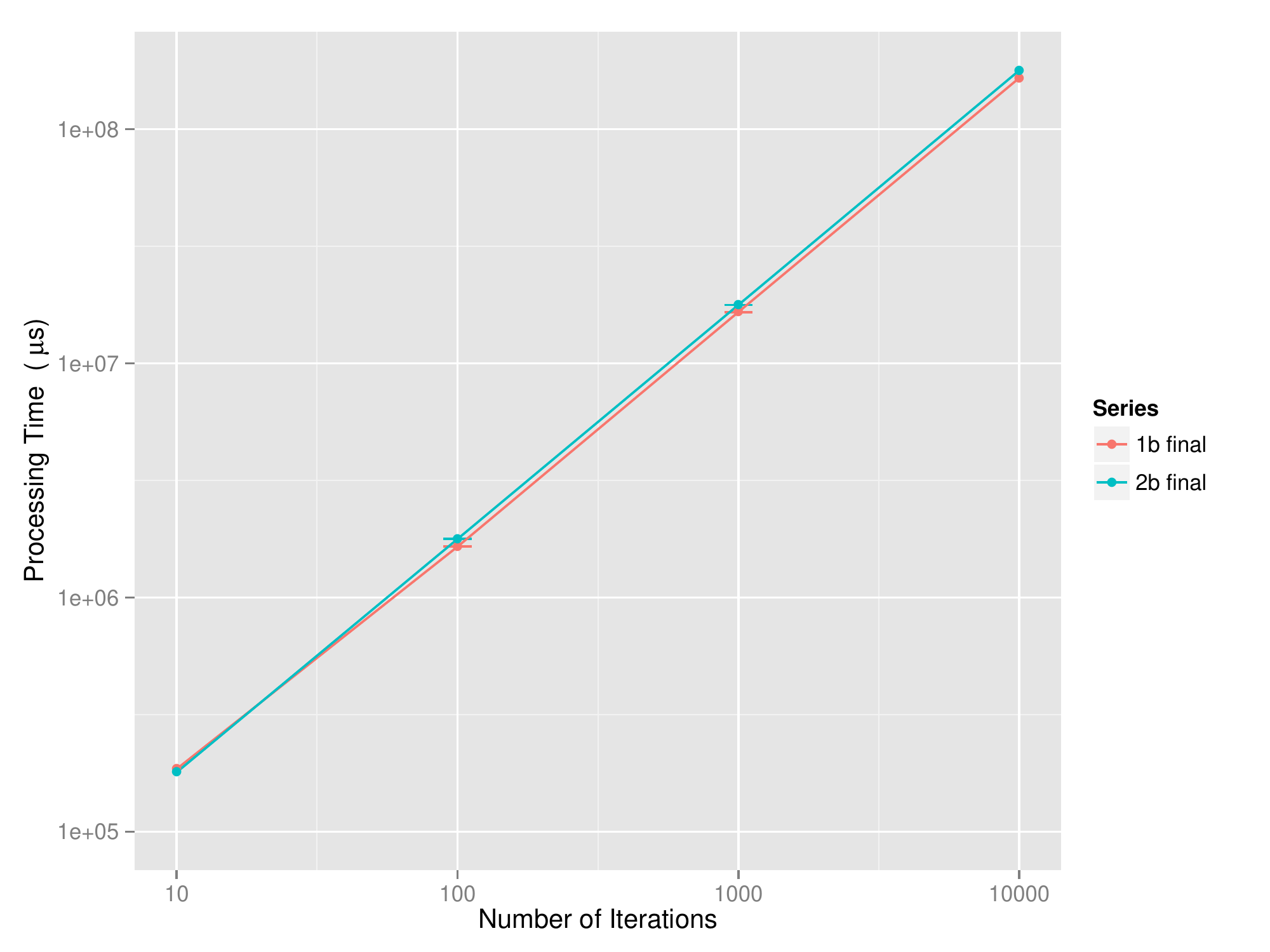}
	\caption{Image Processing Times for \texttildelow8k Image: This graph shows processing times for both the dual buffer (2b final) and single buffer algorithm (1b final) when processing \texttildelow8k images. Although the speed curves appear very close, the dual buffer algorithm is, in fact, slower. Because of the surprising nature of this result, a t-Test was performed, revealing that the difference is statistically significant $(p < 2.2e^{-16})$.}
	\label{fig:graph_sp8k}    
\end{figure}

\begin{figure}[!t]
	\includegraphics[scale=0.45]{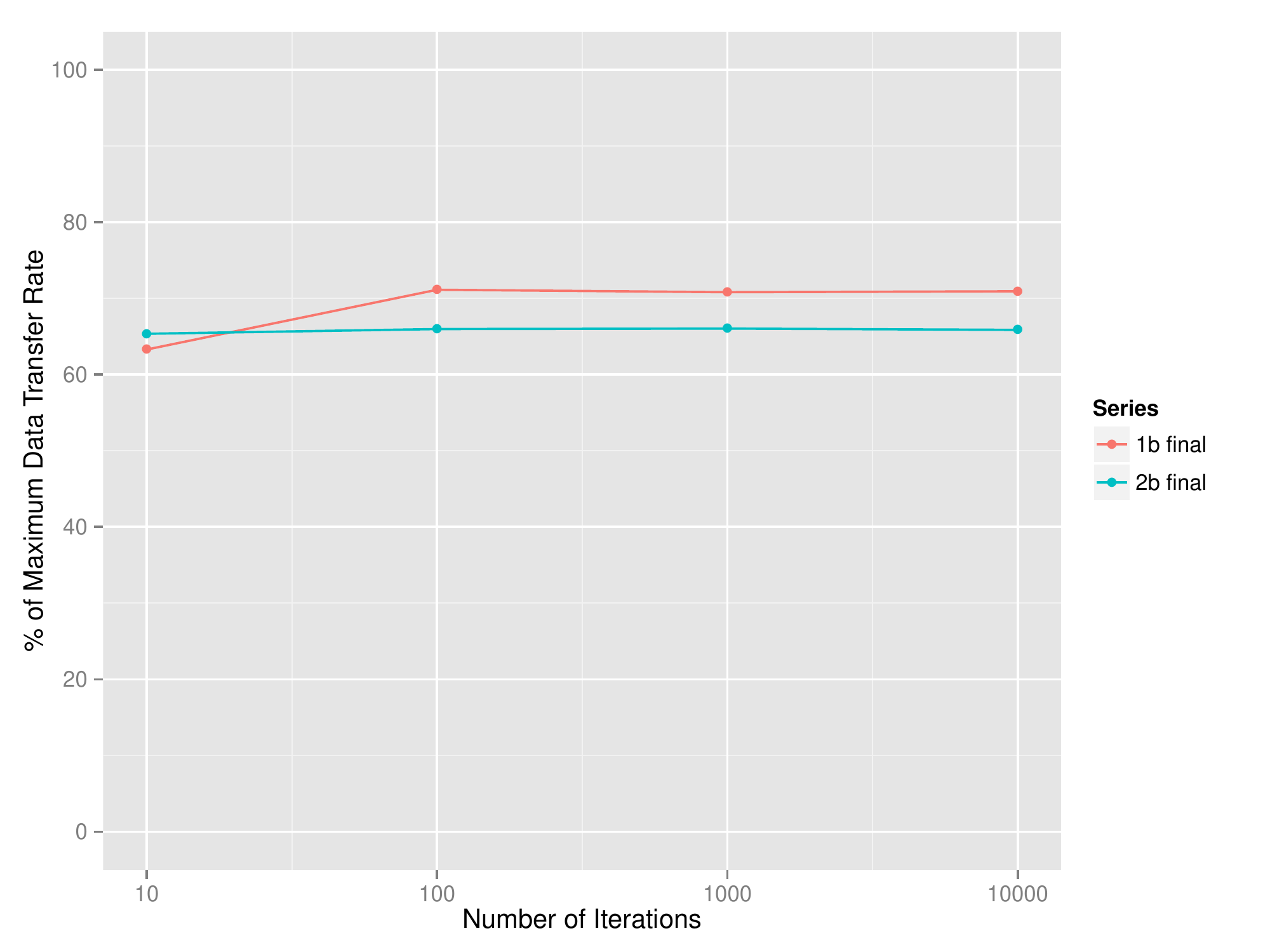}
	\caption{Efficiency for \texttildelow8k Images: This graph shows efficiency for the optimized dual buffer (2b final) and single buffer algorithm (1b final) when processing \texttildelow8k images.       }
	\label{fig:graph_e8k}   
\end{figure}

Tests performed with the \texttildelow8k data set present a very different picture. At first glance, the speed numbers for processing \texttildelow8k images appear to be identical for the optimized one buffer (1b final) and dual buffer algorithms (2b final). A closer examination reveals that 1b final is actually faster than 2b final with \texttildelow8k images, for all values of $N$ except $N=10$ (see Figure \ref{fig:graph_sp8k}). This result is in opposition to all previous tests with datasets of smaller images (see Figure \ref{fig:4ktable}), but the differences are significant based upon a $t$-test ($p\leq 2.2e^{-16}$). With \texttildelow8k image sets, the 1 buffer algorithm in fact demonstrated an efficiency  \texttildelow 5\% higher than the 2 buffer algorithm, for all values of $N$ except $N=10$ (see Figure \ref{fig:graph_e8k} ).

One possible explanation for the degradation of performance with the dual buffer algorithm (2b final) with larger images is that it has higher resource demands, and as a result could strain GPU resources. Essentially, because the two buffer algorithm uses slightly more GPU resources than the single buffer algorithm, it would be the first to experience slow downs at large image sizes. Specifically, the dual buffer algorithm, compared to the one buffer algorithm, requires one additional permanent buffer-image pair. However, dependent on the device manufacturer and model, parts of GPU main memory may not be connected to texture/constant cache structures, making portions of GPU memory unusable for image storage; because available GPU image memory could be less than total GPU memory, the dual buffer algorithm could run into resource contention issues before the one buffer algorithm.

\begin {figure} [!tb]
	\begin{tabular}{ l  l  r  r  r  }
	\hline
	 Algorithm & Data & Iterations & TR   & Efficiency  \\
	    &    Set  &    N   &  GB/s & \%         \\
	\\ [-2.5ex] \hline \\ [-1.5ex]
	1b final & \texttildelow2k & 10 & 2.425 & 51.20\\
	 &  & 100 & 2.647 & 55.79 \\
	 &  & 1,000 & 2.654 & 55.93 \\
	 &  & 10,000 & 2.687 & 56.56 \\
	\\ [-2.5ex] \hline \\ [-1.5ex]
	2b final & \texttildelow2k & 10 &4.073& 85.84 \\
	 &  & 100 & 4.190 & 88.30 \\
	 &  & 1,000 & 4.236 & 89.25 \\
	 &  & 10,000l &4.250 & 89.56\\
	\\ [-2.5ex] \hline \\ [-1.5ex]
	1b final & \texttildelow8k & 10 & 3.536 & 63.29 \\
	 & & 100 & 3.975 & 71.13 \\
	 & & 1,000 & 3.957 & 70.82 \\
	 & & 10,000 & 3.963 & 70.93 \\
	\\ [-2.5ex] \hline \\ [-1.5ex]
	2b final & \texttildelow8k & 10 & 3.652 & 65.36 \\
	 &  & 100 & 3.687 & 65.98  \\
	 &  & 1,000 & 3.690 & 66.04 \\
	 &  & 10,000 & 3.681 & 65.87 \\
	\\ [-2.5ex] \hline \\ [-1.5ex]
	\end{tabular}

	\caption{ 2k and 8k image transfer rate  and efficiency: Efficiency was calculated by comparing the time to transfer and process data, with the indicated algorithms, to the time required for a simple buffer-to-buffer transfer of the same amount of data. The \texttildelow2k image sets has noticeably lower efficiency than the \texttildelow4k which was seen in Figure \ref{fig:4ktrans}. In the \texttildelow8k the one buffer algorithm actually out performs the dual buffer algorithm. }
	\label{fig:2ktrans}
	
\end{figure}

\begin{figure}
	\includegraphics[scale=0.45]{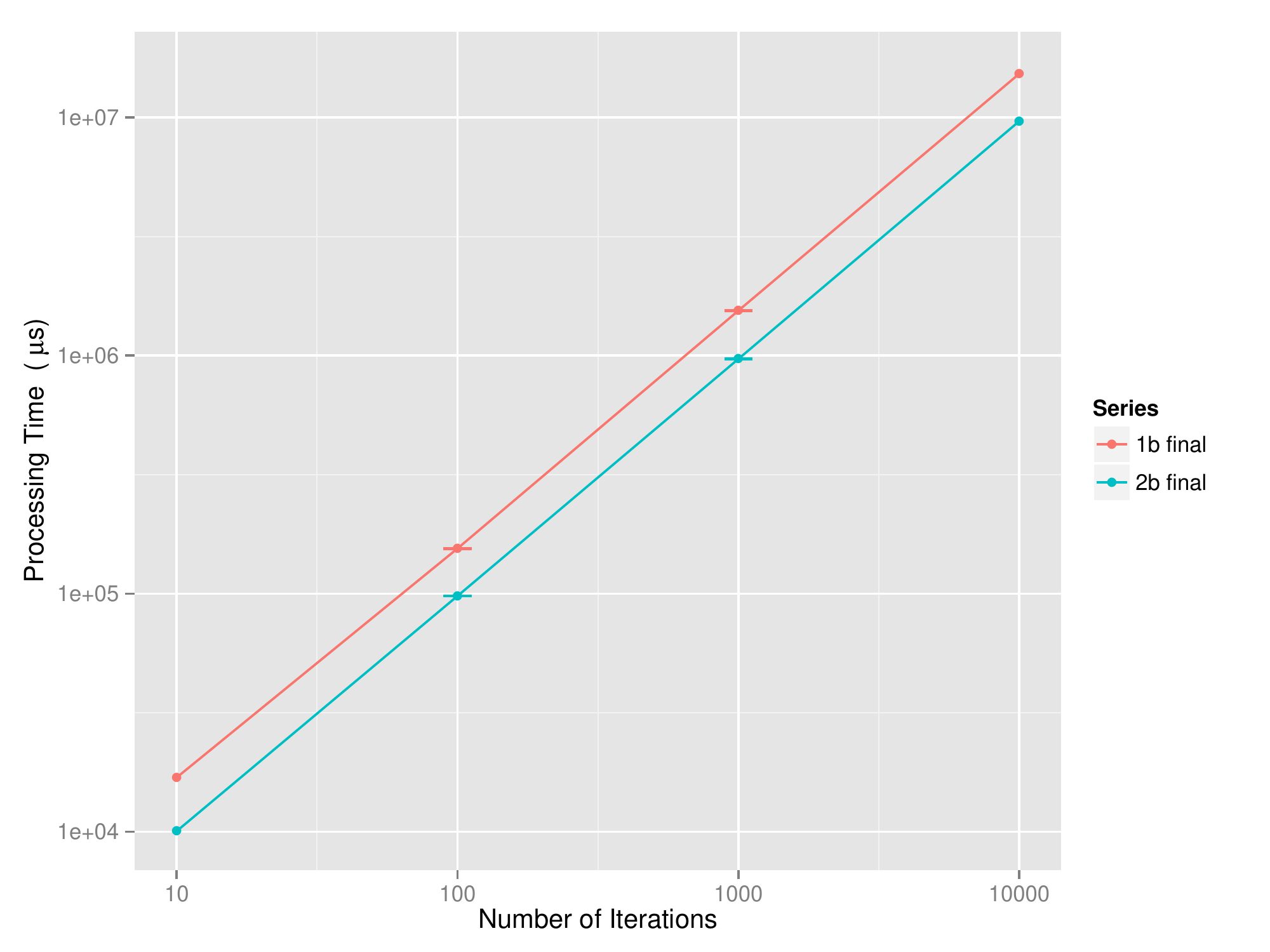}
	\caption{Image Processing Times for \texttildelow2k Images: Parallel lines seen logarithmic space are not actually parallel; in actuality, the timing curves depicted diverge at a near constant rate.  }  
	\label{fig:graph_sp2k}  	 
\end{figure}

\begin{figure}[!t]		
	\includegraphics[scale=0.45]{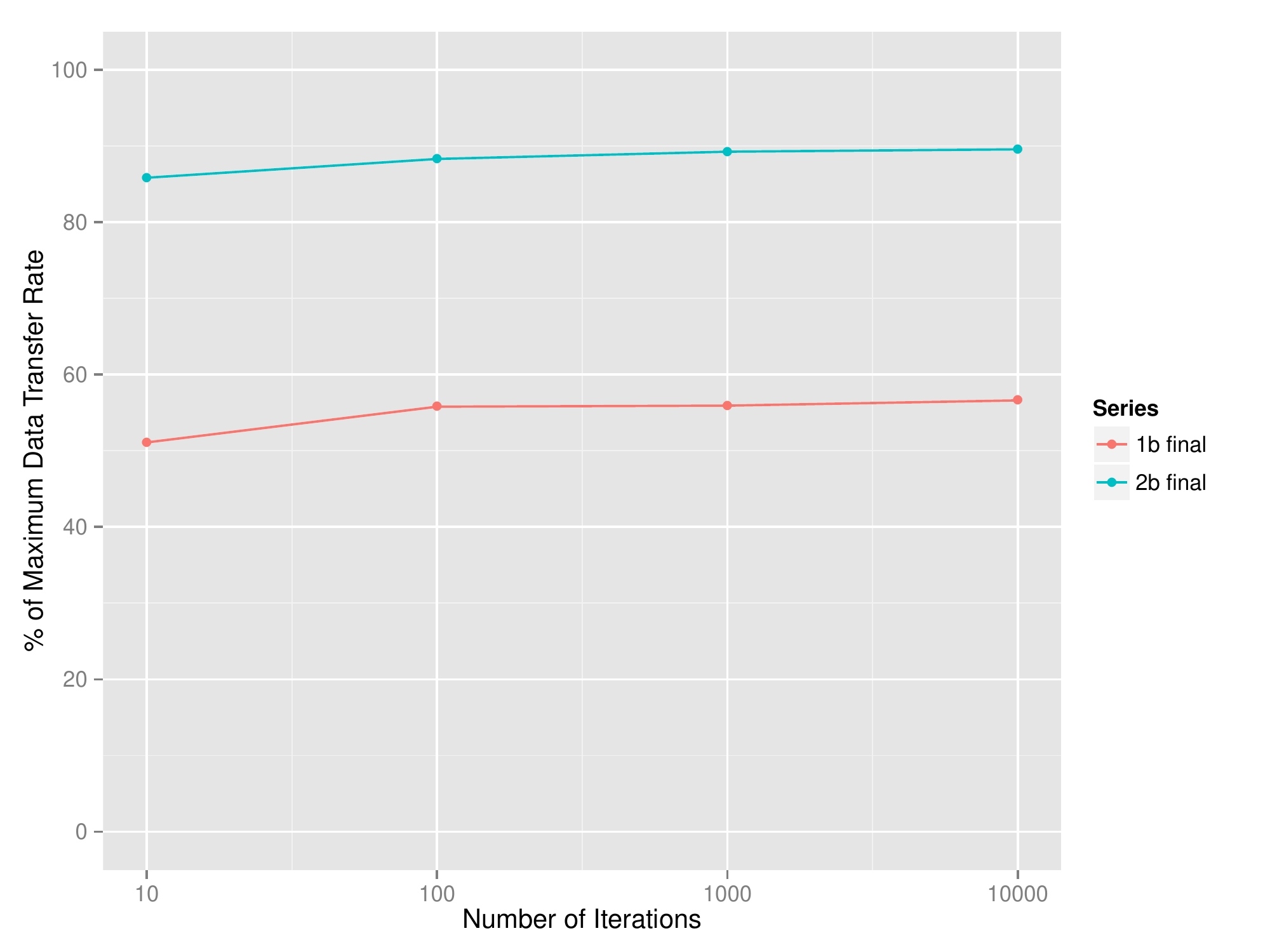}
	\caption{Efficiency for \texttildelow2k Images: efficiency for \texttildelow2k images is noticeably lower than for \texttildelow4k images. The slight decrease in efficiency at N=10 is most likely due to warm up time for the GPU.}
	\label{fig:graph_e2k}     
\end{figure}

\subsection{Buffer to Image Transformation Results}

\begin{figure*}
	\includegraphics[width=75mm]{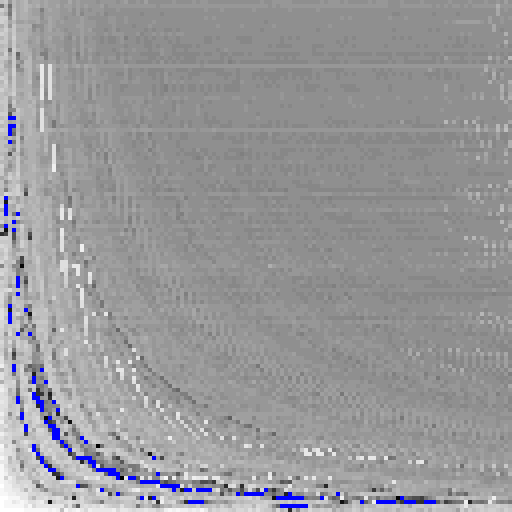}
	\hspace{3 mm}  
	\includegraphics[scale=0.080]{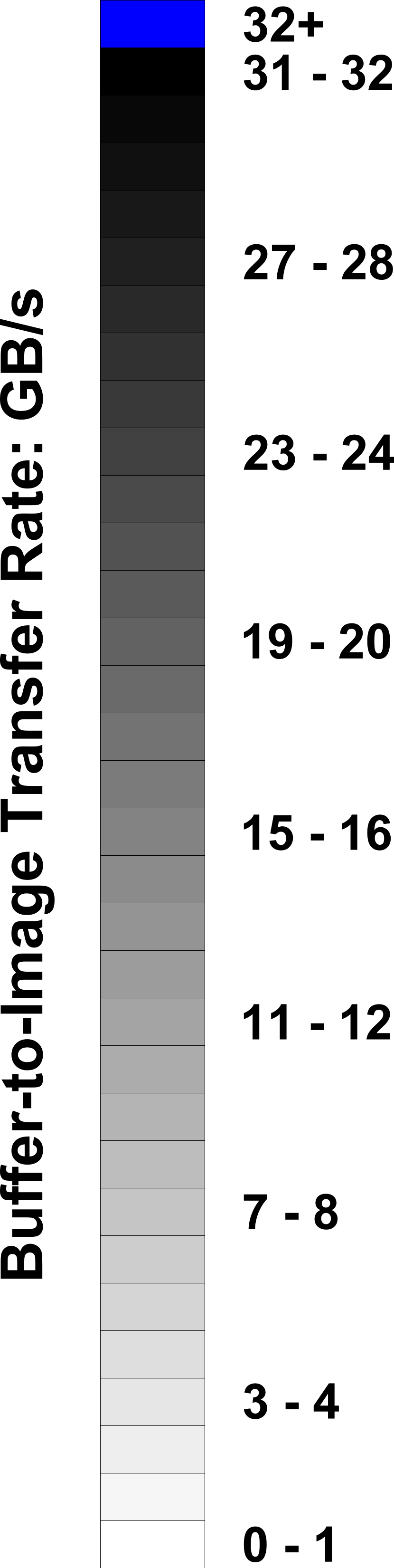}
	\hspace{3 mm}  
	\includegraphics[width=75mm]{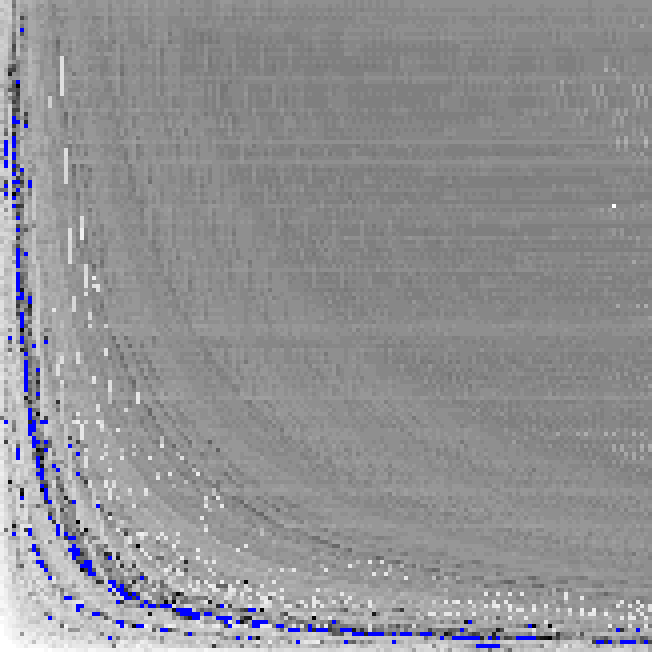}\\
	\hspace*{\fill} (a) 128 size steps \hspace*{\fill}\hspace{20 mm}\hspace*{\fill} (b) 100 size steps \hspace*{\fill}
	\caption{Buffer to Image Transform Transfer Rates: The images above show the transfer rates for all tested image sizes with each pixel representing results for a specific image size. In the left-hand image, the bottom left pixel indicates the transfer rate achieved with128x128 images. Each step right or up indicates an increase by 128 pixels in the X or Y dimensions, respectively; with the upper right pixel representing results for 16,384x16,384 images. The right-hand image follows the same pattern, with the bottom left pixel indicating the results for 100x100 images, and with an X and Y dimension step size of 100 pixels; and with the upper right pixel indicating results for 16,300x16,300 images. The majority of results are displayed with a grey-scale color ramp of 32 steps (with a step size of 1 GB/s) showing transfer rates ranging from white (0-1 GB/s) to black (31-32 GB/s). High outlying results (above 32 GB/s) are displayed in blue. } 
	\label{fig:imagemap}
\end{figure*}

The result of the buffer to image transformation tests can be seen in Figure \ref{fig:imagemap}. The displayed images show the transfer rate recorded when transfer data into increasingly larger images. For example for the 128x128 sampling pattern's image the bottom left pixel shows the results for a 128x128 image, the pixel to the immediately to the right shows the results for a 128x256 image, the pixel immediately above shows the results with a 256x128 image, and so forth. The image showing the results of the 100x100 block sampling works in the same way, the only exception being that image diminsions increase in steps sizes of 100 rather than 128. The images are color coded showing data transfer rate on a grey scale color map. The color white corresponds to transfer rates between 0 and 1 GB/s. The pixel color become dark with each increase in transfer rate in steps of 1 GB/s. The color black corresponds to the transfer rate range of 31 to 32 Gb/s. All transfer values above 32 GB/s are colored blue. Both images display clear bands where the transform rate is much higher than average. Additional bands with less drastic changes in transfer rates can be seen in the larger test image sizes. 

Both images show a banding pattern with curves of alternating high and low performance. The difference between bands becomes less pronounced as overall image size increases. This pattern shows that image dimensions clearly affects the transformation time of images. If mapping had only depended on the amount of data being transferred it should have followed the pattern seen in the buffer transfer tests Fig. \ref{fig:graph_transfer}, and the resulting image should have been retaliative constant, with the exception of very small images. A major implication of this pattern is that use of images can be significantly accelerated by processing data in segments that match one of the higher performance dimensions for the buffer to image transform operation.

 There was no noticeable effect on changing how image size information was passed to image based computation kernels. 

\subsection{Buffer vs Image Kernel Runtime Results} 

The results of the comparison of kernel runtimes when using images and buffers can be seen in Figure \ref{tab:ib}. In the these tests the fastest image kernel was 15.89\% faster than the fastest buffer based kernel. There was no noticeable effect from changing how image dimensions were passed to the image based kernels. The buffer based kernel that used two buffers to handle the accumulation of results noticeably out performed the kernel that used only a single buffer with read and write access. This occurred despite the higher number of memory access operations executed by the two buffer kernel. 

\begin {figure}[!h]
	\begin{tabular}{ l   r  r  }
	Kernel  & Average Time ($\mu$s) & Total Time ($\mu$s) \\
	\hline
	image-kernel 1  & 679 & 6,791,433 \\
	buffer-kernel 1  & 960 & 9,602,299 \\
	image-kernel 2  & 686 & 6,861,058 \\
	buffer-kernel 2  & 807 & 8,074,998 \\
	\hline
	\end{tabular}
	\caption{Kernel Runtimes with Image and Buffer Variants}
	\label{tab:ib}
\end {figure}


\section{Conclusions}

As datasets grow in size, it becomes increasingly important to efficiently process data using GPU memory, especially for applications requiring interactivity. To that end, to appropriately utilize hardware resources, it is necessary to understand the trade-offs inherent in choosing current buffer techniques versus the dual-image approach herein suggested. In this work, we have analyzed the performance of image buffers for scientific visualization kernels and performed a range of tests evaluating their performance standalone and vs.\ traditional buffer approaches. To conclude, we summarize these results and provide some guidelines for the use of image and data buffers in visualization. Future work is also discussed. 

\subsection{Dual Buffer Algorithm}

Based on the observed transfer rates for the hardware tested, 480 MB of data can be processed per frame using the dual buffer algorithm, while maintaining a frame rate of 10 FPS when using 4k images. 2k and 8k images allow 410 MB and 360 MB per frame respectively. These values could be increased by applying compression to the input data and decompressing on the GPU as part of the transformation from buffer to image. Alternatively, increases in the raw transfer rate of the underlying hardware will also increase these values.

An important finding of this research are the scaling factors for the proposed algorithm. The algorithm scaled linearly with increasing iterations for all image sizes tested. For the \texttildelow2k and \texttildelow4k images, the optimized algorithm had better efficiency, whereas the optimized algorithm performed slightly worse with the \texttildelow8k image set. The maximum observed efficiencies with the dual buffer algorithm using the \texttildelow2k, \texttildelow4k, and \texttildelow8k image sets were 89\%, 92\%, and 66\%. Image dimensions were found to have a strong impact on the run time of the algorithm. 

It is also important to note that the performance advantages demonstrated by the dual buffer algorithm do not depend on the the specific computational kernel used in these tests. This is because the observed gains in performance are the result of enabling the system to simultaneously upload data, convert between buffer and image formats, and process images. Whatever processing kernel is utilized has independent run time and will therefore only affect time during that step of the algorithm. This means the dual buffer algorithm could easily be used with an arbitrary computation kernel, for various visualization or computational tasks, while still improving overall efficiency. For example, the dual buffer algorithm could be used to process tiles from a large image where the processing step was the execution of some type of filter; this would not affect the time required to transfer tiles to the GPU, nor the time to convert the tile data from a buffer to an image. The efficiency of the dual buffer algorithm for an arbitrary processing kernel is 
$$  e = max(c_1 + m_{1-n} +p_{1-n}, c_{1-n} + m_n + p_n ) / c_{1-n} $$ 
where $c_i $ is the time need to copy the $i^{th}$ buffer to the device,
$m_i$ is the time to transform the data of the $i^{th}$ image from a buffer to image format, and
$p_i$ is the time required to process the $i^{th}$ image.

One additional finding of this research is that a major cause of performance loss when using external devices is unnecessary synchronization with the CPU. When controlling data processing with the GPU, it is clear that performance can be greatly improved by avoiding synchronization whenever possible. In many cases, the necessary ordering of GPU operations can be achieved by explicitly specifying dependency relationships between events, using event references to control when given events can be executed.

\subsection{Buffer to Image Transformation}

On the tested hardware, a pattern was observed of alternating bands of high and low performance in image dimension space. Knowledge of this pattern can be used to choose image dimensions where performance is high; this allows one of the overheads of using images---the time required to transform the data---to be reduced. Further testing is necessary to see what, if any, pattern exists for other GPUs. In addition, further testing is necessary to see if the observed pattern is maintained when data is being written to part of an image. If this is the case, knowledge of the high performance dimensions could be used to accelerate data transformation into any size image.

\subsection{Buffer vs Kernel Runtimes} 

Using images significantly reduced the runtime of the tested kernel. Based on the simplicity of the tested kernel, it can safely be predicted that use of images will improve the performance of any non-compute-bound kernel. For more precise claims, additional testing with different kernels will be necessary. This test, and the previous test on image transform speed, allow both the cost and the benefit, in terms of run time, for using images as the primary form of I/O to a kernel to be estimated. This allows the correct form of I/O, for optimal performance, to be selected without extensive testing.

Our ongoing work will explore how kernel complexity influences the effect of images vs buffers in computational kernels as well as replication of current tests with Nvidia based hardware to see if the previously discussed patterns hold true. One goal of this testing would be to enable accurate prediction of the effects of changing I/O methods, so that, given a kernel and computational hardware to execute it, an optimal configuration could be created without need for extensive testing.

\bibliographystyle{IEEEtran}
\bibliography{sources}
\end{document}